# A Microscale Nanosecond Time-resolved Platinum Thermometer Probing Gaussian Pulsed Laser Induced Temperature

*Letian Wang*[a], *Dongwoo Paeng*[b], *Zeqing Jin*[a], *He Zhang*[b], *YS. Kim*[b], *Yoonsoo Rho*[a], *Matthew Eliceiri*[a] *and Costas P. Grigoropoulos*[a] **

(a) Laser Thermal Lab
Department of Mechanical Engineering
University of California, Berkeley
Berkeley, CA 94720-1740, USA

(b) Lam Research Corp.
4650 Cushing Pkwy, Fremont, CA 94538

## Abstract:

Pulsed laser processing is playing a crucial role in additive manufacturing and nanomaterial processing. However, probing the transient temperature field during the laser interaction with the processed materials is challenging as it requires both high spatial and temporal resolution. Here, we demonstrate a 9ns rise time 50μm sized Pt thin film sensor for probing the temperature field induced by a nanosecond pulsed laser on a semiconductor thin film. The measurement error sources and associated improvements in the thin film fabrication, sensor patterning and electrical circuitry are discussed. We carried out the first experimental and theoretical analysis of spatial resolution and accuracy for measuring a gaussian pulse on the serpentine structure. Transparent silica and sapphire substrates, as well as 7-45nm insulation layer thicknesses, are compared for sensing accuracy and temporal resolution. Lastly, the measured absolute temperature magnitude is validated through the laser-induced melting of the 40nm thick amorphous silicon film. Preliminary study shows its potential application for probing heat conduction among ultrathin films.

# Introduction

Pulsed laser processing has been widely applied in manufacturing, including machining[1], marking, welding and annealing of semiconductors[2]. Recently it has been incorporated into the additive manufacturing of materials of high melting point and thermal conductivity, including tungsten[3]. Furthermore, pulsed laser processing of nanomaterials has been demonstrated recently, including the sintering of nanoparticles[4,56] and nanowires[7–9] and the directed assembly of photonic structures[10–14]. For a wide spectrum of applications, precise characterization of the involved physical and chemical mechanisms will play a significant role in the understanding of light-matter interaction[15] and improving the manufacturing quality control[16-18]. In-situ characterization at microscale spatial resolution and at nanosecond time scale is therefore desired. Although it provides the most fundamental information for characterizing these processes, temperature measurement at such resolution is still missing, hindering the wide study and application of pulsed laser interaction with materials. The main challenges lie on the $10^{10}$K/s heating and cooling rate, the microscale minimum resolution and the spatial nonuniformity induced by the spatially varying laser beam intensity profiles. It is noteworthy that a Gaussian laser beam is the most widely used beam shape due to its capability for high numerical aperture focusing and stability in long-distance light delivery.

The resistive thermometer is an ideal candidate for probing pulsed laser-induced temperature fields on various materials. In general, the microscale temperature probe can be categorized into three types[19]: electrical, optical and physical contact. Physical contact methods such as scanning probe thermal microscopy or thermocouples typically have slow temporal response due to the large thermal mass of probe. On the other hand, electrical resistance[20,21], optical blackbody radiation[22,23], and thermoreflectance method, feature nanosecond temporal resolution. Though the non-intrusive nature of radiation and thermoreflectance offers unique strength, mounting a bulky optical apparatus onto the laser processing setup is both time- and resource- consuming for R&D. Another limit of the optical method is that it is not capable of operating with the broadband energy sources such as lamps and plasma, commonly used in semiconductor processing, because they will interfere with the probing optical wavelengths. Furthermore, surface modifications such as ablation, melting, or oxidation[24] introduced by the laser will also affect the optical response, lowering the accuracy of the temperature measurement. Electrical resistance is ideal for studying the laser material interaction as it does not require a dedicated optical setup, nor prior optical property characterization and analysis of the sample surface condition.

High spatial and temporal resolution temperature measurements of Gaussian laser pulse irradiation are absent from the literature. Brunco et al.[20,21] developed a Pt thin film resistive thermometer to probe the nanosecond laser heating of thin silicon films. A Pt film was embedded underneath the absorbing layer with nitride as the insulation layer. The sensor identified the melting temperature for pure Si and its alloy assisted by heat transfer simulation. However, the rise time was found to be 100ns, and the spatial resolution is also limited to 1mm. A dynamic thin-film based microscale thermocouple[25] with 28ns rise time and 25 μm size has been fabricated. However, the sensor does not probe the temperature field of the laser irradiated area. Instead, it probed the region proximal to the laser irradiated area. Furthermore, none of the sensors demonstrate the capability to probe temperature differences caused by the subtle variation of film thicknesses. Lastly, all the above measurements did not address Gaussian shaped beam profiles associated with focused laser beams. A systematic analysis of the required sensor pattern and algorithm adjustment based on the thermal analysis should be carried out.

In the current study, we show the design and validation of a thin film Pt sensor of 9 ns rise time and 50μm active area. We demonstrate the capability to detect temperature distributions induced by Gaussian laser beam profiles, discuss the error sources and corresponding optimization steps before establishing its spatial, temporal resolution, accuracy and precision. Coupled thermal and electrical simulation is employed for comprehensive theoretical analysis. Furthermore, we confirm the absolute level of sensor measurement against nanosecond laser-induced melting of an amorphous silicon film. Lastly, we discuss the sensor's applications for probing the heat transfer of multilayer ultrathin films.

## Sensing Instrument Design and Implementation

We have demonstrated the fabrication and calibration of a 50 μm sized 50nm thick Pt sensor on transparent substrates [26] ideally designed for probing transient temperature fields induced by pulsed lasers. The insulated sensor structure is illustrated in Fig. 1a. The same sensors and Temperature Coefficients of Resistance (TCR) are used in the current study. The sensor is laser machined into a serpentine structure and its electrical contacts are directly connected to the lead wire with silver paste (Fig.1a.). For laser processing, an absorbing layer will be deposited on top of the sensor and receive laser energy. Our previously reported work[26] showed that the sensor capped with an additional germanium absorbing layer can provide repeatable TCR performance over several cycles of heating and cooling. For characterization of the sensor performance, both sensors with and without top layers are studied.

Fig. 1b shows the schematics of the sensor's working principle. The sensor temperature after laser irradiation forms a T(x, y, t) field because the Pt layer is thermally thin in the z-direction. The temperature distribution T(x, y, t) leads to the electrical conductance distribution σ(x, y, t) due to σ(T) correlation. Different from Brunco et al[20] flat beam irradiation, the current sensor has a serpentine structure and is expected to probe Gaussian beam irradiation. As a result, the detailed temperature distribution T(x, y, z, t) and associated σ(x, y, t) will be further discussed in later sections. Note that interested RF signal characteristic wavelength can be estimated to be 3 to 30m, based on $\lambda = \frac{c}{f}VF$ and a velocity factor(VF) of 0.5 as well as a frequency f is around 10-100MHz. Since the sensor size is 2 to 5 orders smaller than the RF signal wavelength, we infer that the local variation σ(x, y, t) will not affect the RF signals beyond a lumped resistance R(t). The R(t) then affects the measured V(t) through the customized circuit. Based on the transient V(t) and other static voltage measurement, we can formulate the probed resistance $R_{prob}(t)$ into algorithms. Lastly, the resistance is translated to the probed temperature $T_{prob}(t)$ based on calibrated TCR from ref[26].

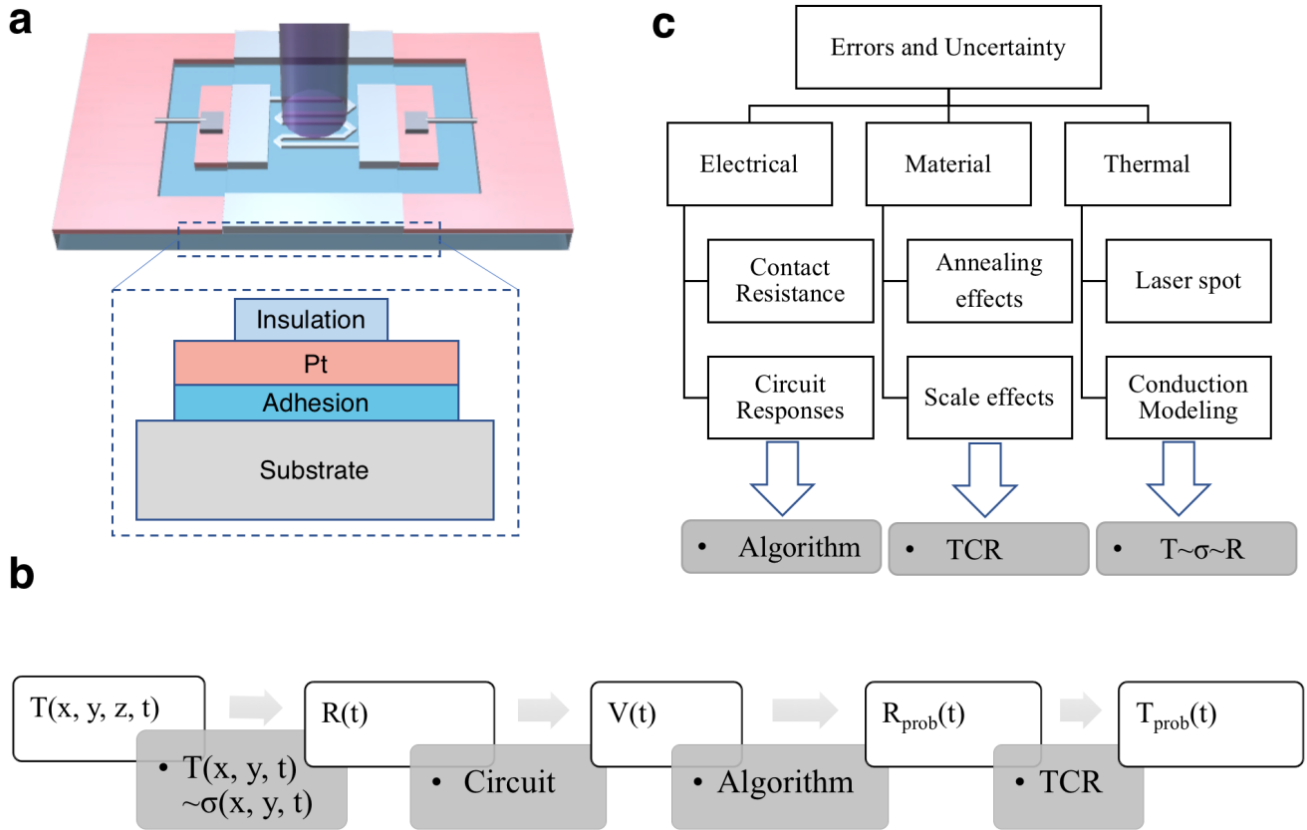

**Figure 1 Schematics of the sensor, its working principle and error analysis.** (**a**) The schematics of the sensor with an inset showing its cross-section. (**b**) the working principles with upper row showing the data flow and the lower row showing the correlations, (**c**) the error analysis

As depicted in Fig. 1c, the sources of errors are categorized into three fields, i.e., the electrical, thermal and material. Electrical errors are associated with the contact resistance and circuit response. Material errors comprise the sample resistance stability against different thermal annealing effects from fabrication and the degradation effects associated with laser machining. Thermal errors are associated with laser spot size and the serpentine sensor structure. Based on these analyses, we illustrate the efforts on sensor design and fabrication towards reducing error and improving sensitivities in the following sections.

## 1. Sensor Fabrication

The substrate substantially affects the laser-induced spatial and temporal temperature evolution. As mentioned in our recent work[26], the transparent substrate allows for heating from top or bottom. Sapphire and silica are two chemically inert and thermally stable substrates with melting points reaching 1617 ºC and 2050 ºC, respectively, higher than most metals or semiconductors. From the thermal conductivity point of view, these substrates have distinctly different thermal conductivities (silica is 1.5 W/mK and sapphire is 25 W/mK). The selection of substrate offered two typical cases for real substrates at the time scale we are interested in. For ultrafast laser processing, most thermal processes conclude within 100ns. The thermal penetration depth in both aforementioned materials is within 1μm. Consequently, silica substrates resemble applications where the Pt layer is backed by silicon oxide or other amorphous materials. On the other hand, sapphire mimics the thermal behavior of substrates having high thermal conductivities, such as crystalline Si, Ge, and III-V at elevated temperatures. To improve the adhesion of Pt film to silica, we deposited 7nm $Al_2O_3$ through atomic layer deposition (ALD) as an adhesion layer[26].

Depositing an insulation layer between the Pt film and the probed specimen is critical as it prevents electrical shortage, chemical diffusion or mechanical damage to the Pt layer. The sensor film quality is affected by the thickness of insulation layer. In Table 1, the Pt film is insulated with 45nm of PECVD oxide (350 ºC, 15min) or 7nm thick alumina (300 ºC 40min). Sheet resistances are reduced to a similar level after depositing different insulation layer. The main cause is the high-temperature annealing accompanying the deposition processes. Insulation can also improve the mechanical properties of the film from brittle to ductile with the enhancement of adhesion (Supplementary Information Fig. S1). The sensor's TCR performance is adapted from the cooling parts reported in ref.[26] The thickness of the insulation layer will further affect the sensor's sensitivity and its potential application is discussed later.

Table 1 Sheet resistance measurements of Pt film on different substrates before and after depositing different insulation layers

| No. | Substrates | Adhesion Layers | Before Insulation (Ω/sq) | Insulation with Oxide (Ω/sq) | Insulation with Alumina (Ω/sq) |
|---|---|---|---|---|---|
| 1 | Silica | - | 3.81 | 2.58 | 2.57 |
| 2 | Silica | $Al_2O_3$ | 3.46 | 2.67 | 2.54 |
| 3 | Sapphire | - | 3.37 | 2.70 | 2.71 |

## 2. Electrical Circuit Design and Signal Processing

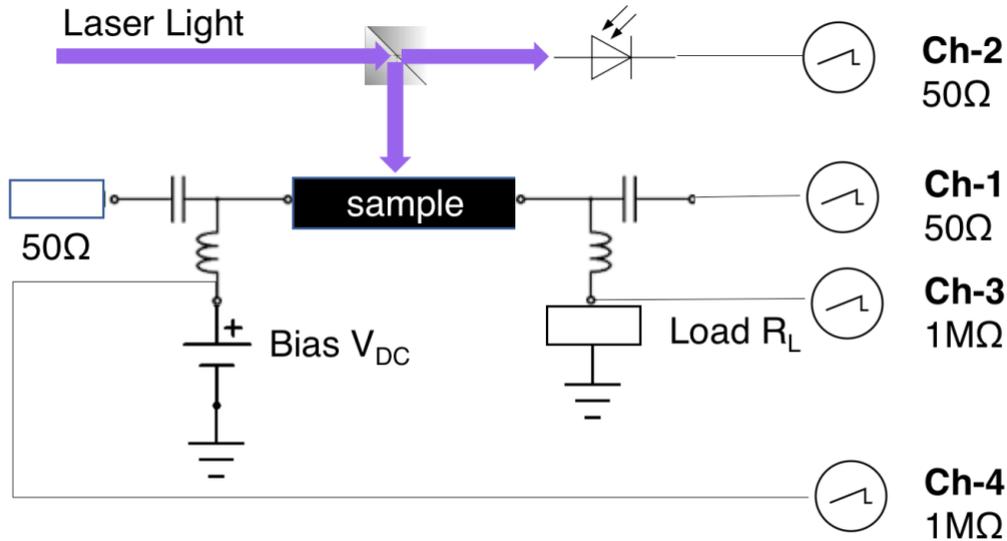

**Figure 2 Schematics for the electrical circuits and experimental setup.** Channel 1-4 records transient voltage, laser triggered photodetector signal, DC bias voltage and DC load voltage.

Different from the " transmission line" setup by Brunco et al.[21], a clear DC and RF separated path is realized through 2 identical bias tees (Fig.2), which eliminated the tuning and matching of the capacitor and inductor. The DC route biases the sensor and constitutes the charge conservation to formulate the original resistance $R_0$. The RF route is dedicated to isolating the transient voltage caused by the resistance change. The bandwidth of the RF is greater than 4GHz based on the specifications of bias tee. Four channels of signals are terminated in a 500MHz 4G sample/s oscilloscope (Rigol, DS4000). A LabVIEW program is developed to simultaneously record the signals and interpret temperature *in situ* with laser processing.

The error associated with the circuit can be recognized mainly in the waveform distortion and the response time limit. The waveform distortion is caused by impedance mismatch and is discussed in this section, while response time limit will be discussed at the temporal resolution section. All the electrical connections in the RF route are made with coaxial cables that match the impedance of 50 Ohm. The coaxial optimized probe station cannot fit into the chambers where laser processing happens. Therefore, the parasitic capacitance and inductance of our lead wires will serve as sources for impedance mismatch. The analysis of captured waveform entails a ringing frequency of 50 MHz (Fig. 3a, iii), whose effective wavelength is estimated as 6m based on VF*c/f and velocity factor (VF) of 0.5. The clip and leads are 1m in total length, two orders larger than the dimension of the sensor (<1cm) and less than one order smaller than the effective wavelength. Therefore, we confirm the clip and leads are the main sources of impedance mismatch and the resulting ringing. The mismatch can be reduced by registering the sensor to an impedance matched PCB board through wire-bonding, which is not pursued in this study.

Finite Impulse Response (FIR) digital filters are compared for ripple signal removal. In Fig.3 we show the frequency domain optimized low-pass filter can preserve better peak features than time-domain optimized Savitzky-Golay(S-G) filter. It is shown that the S-G filter can be tuned towards the performance of a low pass filter at a certain time scale (Fig. 3bi), but failed at another (Fig. 3bii). It is mainly because the ringing signals have a characteristic frequency at 50MHz which can be easily removed by a low-pass filter. In Fig. 3cii, the frequency ripple has been removed for filter

type=2 at a frequency beyond 50 MHz. However, even with an optimized filter, we still lost the granular peak information for the signals of Fig. 3ai. Larger error on the maximum temperature is generated on a slower heating and cooling process (Fig. 3b compared to 3a). As a consequence, no digital signal processing is adapted for comparing measurement with simulations.

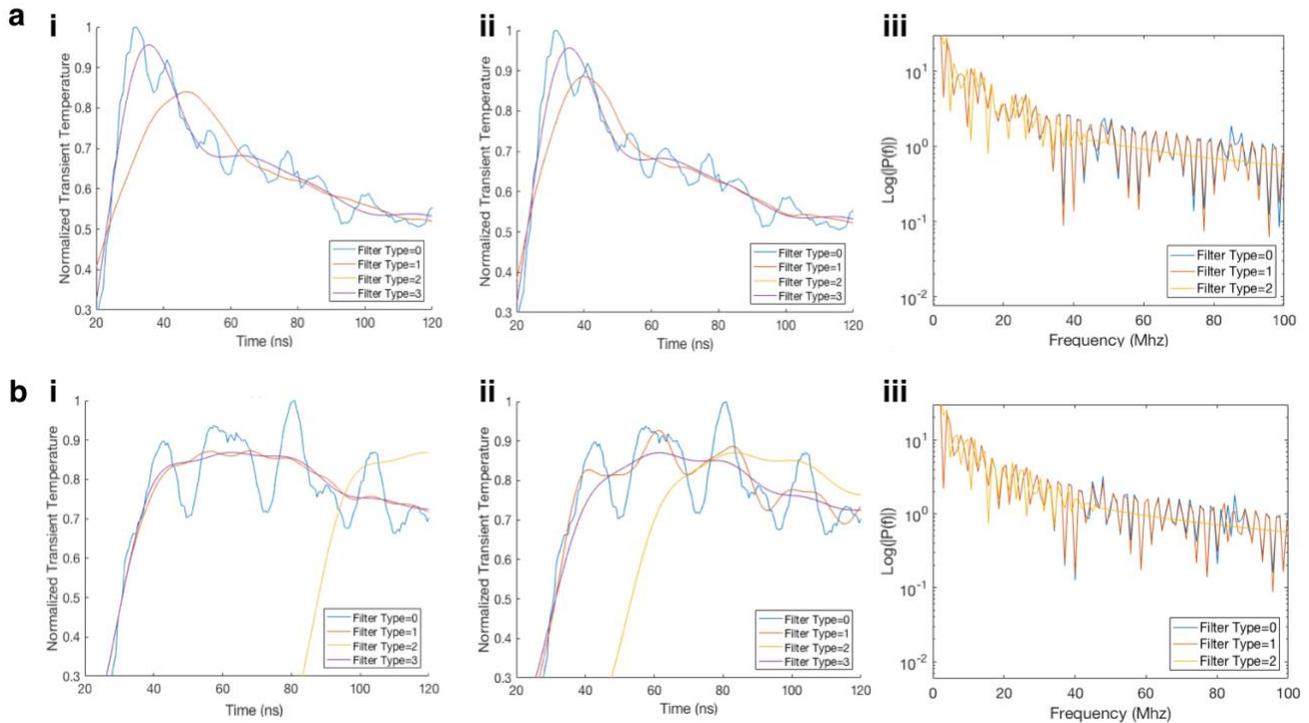

**Figure 3 Ringing effect and comparison of signal processing options.** Signals of nanosecond laser irradiation on the sample with 50nm Ge deposited on (a) 7nm alumina as an insulation layer or (b) 45nm oxide as an insulation layer. (i-ii) Comparison of low pass filter with S-G filter with (i) a large number of elements or (ii) a small number of elements; (iii) the FFT based frequency analysis of signal after filtering. In the legends, the filter type "0" stands for no filter, "1" is Savitzky-Golay filter with an element size of 73 or 53 and order of 3, "2" is low pass filter with an element size of 88, cut-off frequency 100MHz, and "3" is the time-corrected "2".

## 3. Serpentine Sensor Pattern

The serpentine structure was suggested but not discussed in any details in Brunco's paper[21]. In general, the serpentine structure can effectively increase the Pt sensing unit resistance and reduce the contact resistance error while keeping a compact footprint. However, it generates thermal error under Gaussian beam irradiation, which will be discussed in the spatial resolution and accuracy section.

Illustrated in Fig. 4, probe resistance and contact pad resistance are bundled series resistances in the circuit. Note the $R_{TC}$ here includes the contact resistance between the probe and contact pad, termed as $R_C$, as well as the contact pad intrinsic resistance $R_{CP}$. Through silver paste bonding, we reduced the contact resistance $R_C$ to the minimum value of below 0.5 Ohm. The four-point probe method is not applied here due to the circuit design illustrated in the above sections. The contact pad intrinsic resistance $R_{CP}$ is defined by the space required to implement silver paste bonding, which contributed to the majority of $R_{TC}$. Therefore, the probe resistance $R_T$ should be engineered significantly higher than $R_{CP}$ with a given lateral dimension. According to Ohm's law $R = \frac{\rho \cdot L}{A} = \frac{\rho \cdot L}{W \ast d}$, the ratio of $\frac{L}{W}$ determines the resistivity in a thin film pattern, where d is the thickness, and L and W are the total length and width of the pattern. The silver paste is controlled to be 4mm wide($W_{CP}$=4mm) and the distance to the sensor pattern is controlled to be 1mm($L_{CP}$=4mm). As $\frac{L_{CP}}{W_{CP}}$ is 0.25, the $\frac{L_T}{W_T}$ ratio is then required to be at least 20. From the densely packed serpentine structure, we found an N fold serpentine structure has $\frac{L_{tot}}{W_{tot}} = \frac{L_{ind} \ast N}{W_{ind}}$, where L and W are the length and width, the subscript "ind" and "tot" stand for an individual stripe or the total pattern inside the serpentine structure. The definitions can be further referred in the Fig. 4. The larger $W_{tot}$ is preferred to reduce the thermal error. The lower bound of $W_{tot}$ is set to avoid laser machining induced defects and surface agglomeration in the vicinity of the edge, requiring $W_{tot}$ larger than twice the femtosecond laser beam diameter used for machining. The problem is then transformed to find maximum $W_T$ with constrains as follows : $2*D \leq W_{tot} \leq L_{ind}/N$, $\frac{L_{tot}}{W_{tot}} > 20$, D is the diameter of the laser machined spot, which is 1 µm. It is easy to see the optimal solutions lies on the boundary where $W_{tot}=L_{ind}/N$, so the $\frac{L_{tot}}{W_{tot}} = N^2$. Therefore, the maximum $W_{tot}$ is achieved at minimum N=5. The typical resistance of a five-fold serpentine pattern on 50 nm Pt film after annealing is measured to be 60-70Ohm. The residual resistance is measured to be 1-2 Ohm. Therefore the contact resistance error is controlled below 2%.

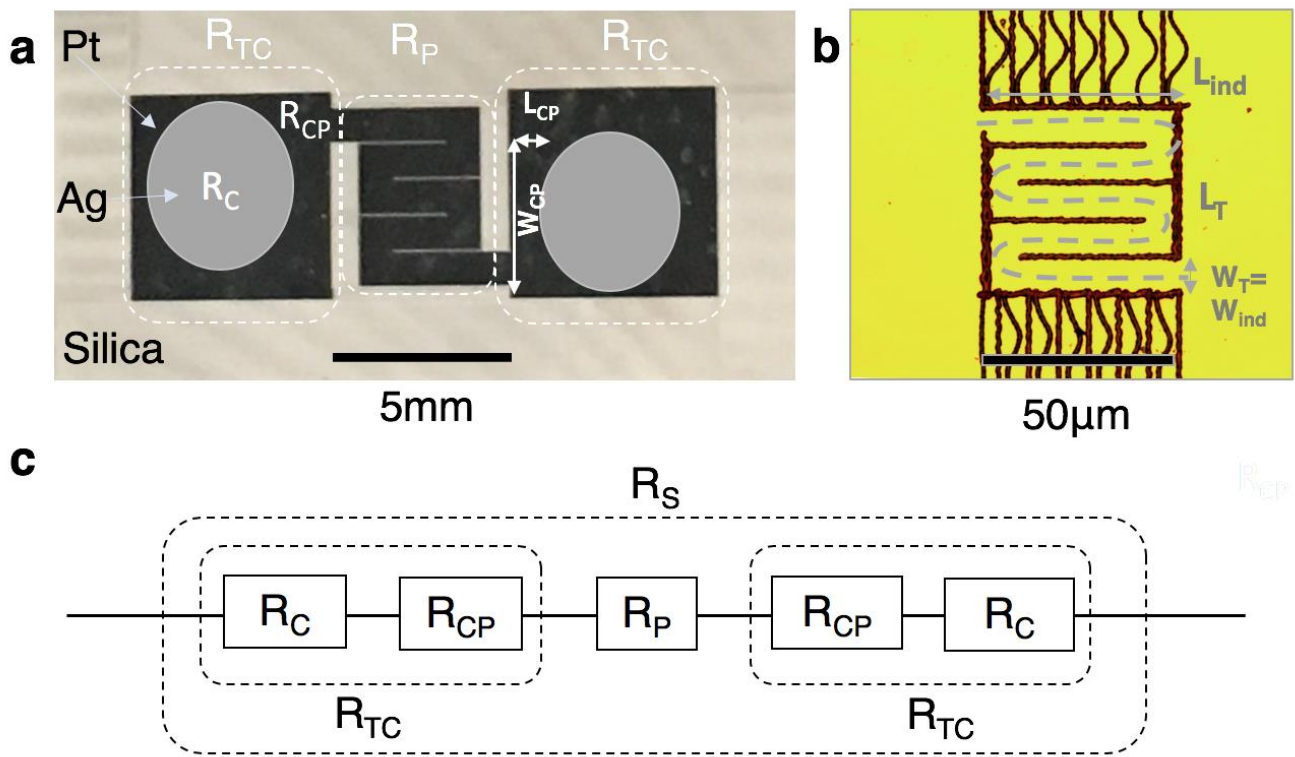

**Figure 4 Laser machined sensor geometries and electrical resistance analysis**. (**a**) The image of a 5-fold serpentine structure with 4mm active size. (**b**) Microscopic image of a sensor with a 50μm active area. The dimensions of $L_{ind}$, $L_{tot}$ and $W_{ind}$, $W_{tot}$ are labeled respect to the sensor. (**c**) The breakdown schematics of the resistances of the sensor. $R_S$ is the sensor resistance, and $R_T$ is the probe resistance. $R_{TC}$ is the total contact resistance including the $R_{CP}$ the contact pad resistance and $R_C$ the contact resistance between the probe and contact pad.

## Sensor Performance and Applications

We analyze and describe the three performance indicators, i.e., resolution, accuracy, and precision. Our analysis will examine both the maximum temperature and normalized temporal variations. For laser-induced damage, melting or deformation threshold studies, the maximum temperature is of interest. For thermal conductivity or frequency domain measurement, the temporal evolution is more important than the maximum temperature. On the other hand, for probing laser-induced chemical reaction processes, both the maximum temperature and temporal evolution should be considered.

Before our discussion, we first analyze the heat transfer problem involved in the sensing process. The vertical thermal diffusion length $L_{\text{diff}} = \sqrt{\alpha t}$ is below 500nm for 10ns and 2.5μm for 250ns for both oxide and sapphire substrates. The full-width half maximum (FWHM) of the laser-irradiated area is 50μm. Therefore, the vertical temperature gradient $T_{\text{peak}}/L_{\text{diff}}$ is by 1-2 orders larger than the lateral gradient $T_{\text{peak}}/(1/2\text{FWHM})$. Hence, the 1D vertical heat dissipation to the substrate dominates the temperature evolution. For the actual sensing processes, however, the measured temperature represents the lumped thermal resistive effects over the entire sensor. Additional details are revealed from coupled thermal and electrical simulations.

## 1. Spatial Resolution and Accuracy

In Fig. 5, coupled heat transfer and electrical conductance simulation is carried out in COMSOL with a simplified 3D transient model. Insulated with a 7nm ALD alumina layer, a 50μm wide 50nm thick Pt sensor on sapphire is irradiated with a 13 nanosecond laser pulse with different laser beam diameters. The laser fluence is set to generate a temperature change of 500K to 600K, which is the temperature range of interest. For small beam sizes (FWHM=25μm, Fig. 5ai), we note the temperature field on the sensor shares a similar Gaussian distribution with the incident laser beam.

The spatially averaged temperature increase is only half of the center temperature (Fig. 5bi). On the other hand, the large laser beam (FWHM =100μm) irradiation induces a uniform temperature distribution on the sensor pattern. As a result, the average temperature is only 50K lower than the peak temperature increase (500K). The 10% difference between the average and the center temperature is caused by the cooling of the sensor's edges. In the zoomed-in image and cross-section temperature plot from Supplementary Information Fig. S2, the edge temperature has dropped to half of the center temperature. The cross-sectional average temperature is consistent with the overall temperature. Here we define a "Resistive T" representing the calculated probed temperature $T_P$ based on the $\frac{R_P}{R_0} = TCR * (\frac{T_P}{T_0})$, where $R_P$ is the simulated overall resistance change. From the plot in Fig.5b, the "resistive T" is identical to the average T regardless of the temperature distribution, which indicates the actual probed temperature will be the spatial average temperature. Further explanation on this identity is not pursued due to the scope of current work. In summary, the spatial resolution of the sensor is 100μm and the system error is 10% due to the effect of serpentine structure and Gaussian irradiation.

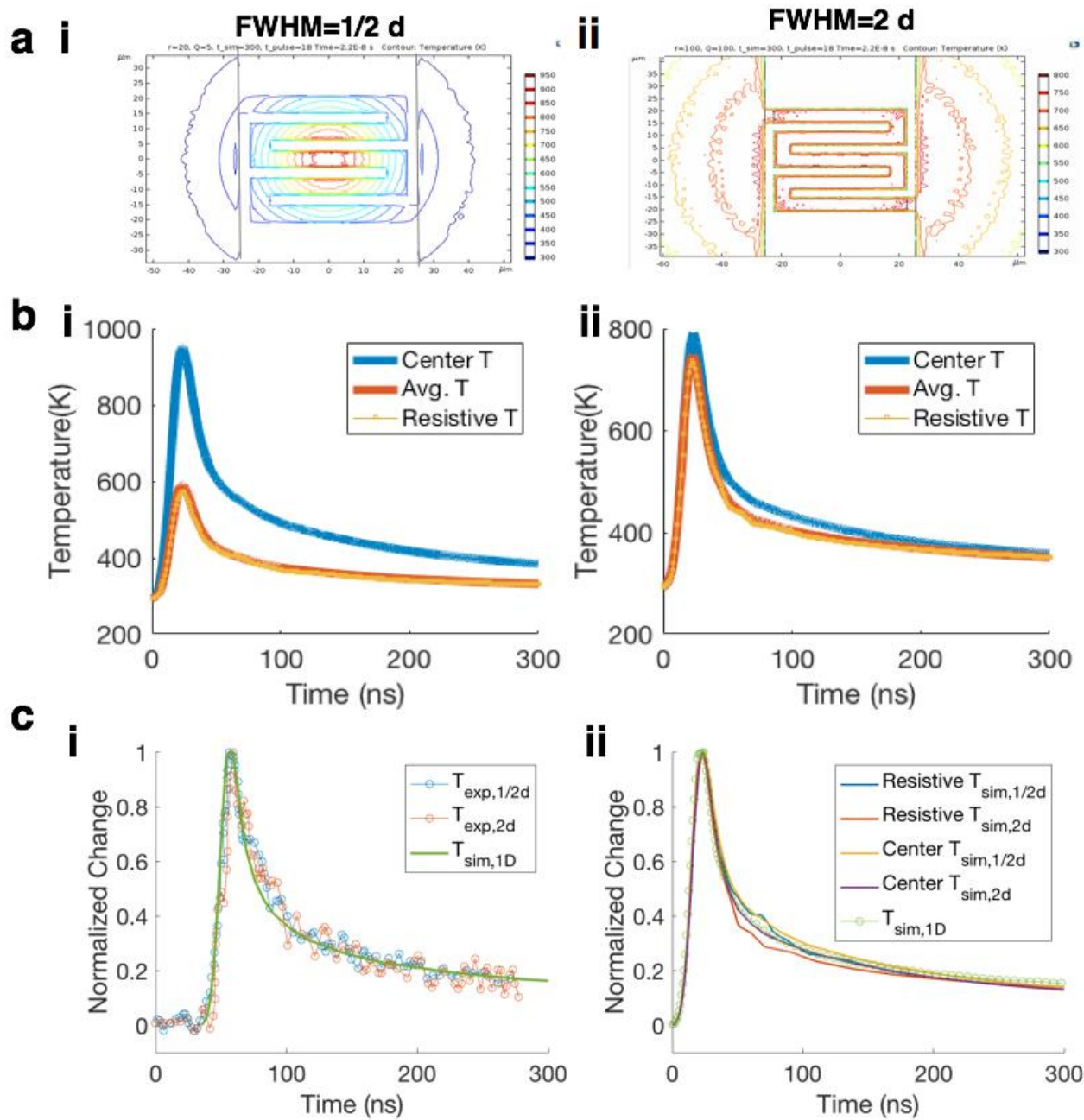

**Figure 5 Experiments and simulation for the spatial resolution and associated thermal error for 50μm size sensor on sapphire.** (**a**) Simulated temperature contour of the sensor exposed to a 13ns laser pulses with beam sizes of (i) 25μm diameter and (ii)100 μm diameter. Here d stands for the width of the sensor, which is 50 μm. The plot is selected at the 22ns when the peak temperature is achieved. (**b**) The time evolution of the simulated center temperature ("Center T"), average temperature ("Average T"), and resistance back-calculated temperature ("Resistive T") under laser beam of (i)25μm diameter and(ii)

100 µm diameter. (**c**) (i) the experimental temperature evolution history under two sizes of laser beams. (ii) The simulated center T and resistive T evolution under two sizes of laser beams. All the samples are insulated with 7nm ALD alumina without a top absorbing layer.

The edge-induced error can be reduced or properly accounted for through post-processing. Expanding the stripe width can reduce the ratio of cooled edge width to stripe width. However, for a given 50µm sensor size, stripe width has a maximum value of 7.5µm for a five-fold serpentine pattern, limiting the minimum error. Alternatively, the oxide substrate can reduce the edge cooling as it has one order lower thermal diffusivity compared to sapphire. It is noteworthy that the edge cooling induced error is a systematic one that can be properly estimated with COMSOL simulation and corrected as a ratio regardless of laser fluences. Post-processing correction is a viable route for removing the error.

We further show that the temperatures normalized with respect to the peak signal value are dominated by one-dimensional (1D) vertical heat dissipation to the substrate. These normalized transients agree well with the 1D simulated temperature evolution in Fig. 5ci. Furthermore, in simulations presented in Fig. 5cii show that the heat dissipation from the edge of Pt stripe only affects the temperature evolution at the transition from the initial fast drop to the long-tail cooling. The comparisons conclude that with proper accounting for the edge effects on maximum temperature, the interpretation of the transient signal can be based on the 1D vertical heat transfer.

## 2. Temporal Resolution and Accuracy

Besides the spatial resolution, we will also experimentally analyze the temporal resolution and accuracy. The temporal resolution is defined by the response time, which is characterized by evaluating the impulse response of the sensor. Picosecond laser pulses are by two orders faster than the thermal relaxation time (~1ns) as well as the oscilloscope response time (1ns) and they are considered as impulse inputs into the system. The sensor's response is listed in Fig. 6a. We found the signal first increased instantly, then gradually reached a peak within 12 ns and started dropping afterward. From basic 1D heat transfer analysis, the temperature increase inside the Pt layer should conclude within 1ns. However, it takes 12 ns to reach the peak, which invalidated the measurement of the peak temperature. We think the main cause of such a long response time (9ns, 1/e) is on the non-ideal DC power supply[21]. The charge cannot restore within 1ns. Therefore, voltage first increases to a medium level, then continues to increase gradually while the charge is being restored. Asymptotic curve fitting (dashed line in Fig. 6a) may help to estimate the peak temperature; however, the error will be significant as the curvature near the peak is high.

For the temporal accuracy, we compared the experimental and simulated sensor's normalized temperature evolution on different substrates. In Fig. 6b, sensor response signals on both sapphire and oxide matched well with the simulation. Since the sensor's maximum temperature is reached beyond the response time (shadowed area in Fig.6b), no significant distortion shall exist for the probed peak temperature and following cooling processes.

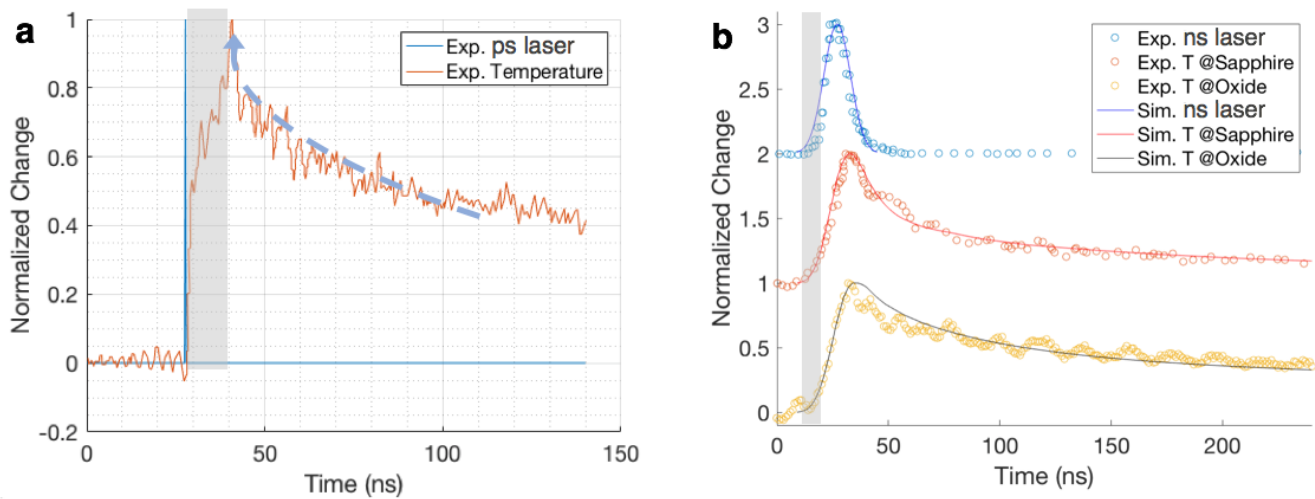

**Figure 6 Temporal evolution upon picosecond and nanosecond pulsed laser irradiation.** (**a**) Experimental laser signal and sensor response upon picosecond pulse irradiation on an oxide substrate. (**b**) Both experimental and simulated laser signals and sensor responses upon nanosecond laser irradiation on sapphire and oxide substrates. All the samples are insulated with 7nm ALD alumina.

## 3. Measurement Precision

Only the temporal precision, both magnitude and evolution, are discussed. Since sensor has a fixed location, the spatial precision of measurement only depend on the imposed laser beam intensity distribution. The laser energy spatial distribution fluctuation induced precision loss is lumped into the temporal precision of the magnitude. Ten independent measurements are carried out through shining a continuous train of 13 nanosecond laser pulses onto a Pt sensor with 45nm oxide insulation and backed by a silica substrate. We first confirmed the temporal alignment of the temperature evolution through cross-correlation evaluation of all the signals. Then we note the measured the transient peak temperature increase has an RMS level of 197.27 K and an RMS error of 7.24 (Fig.7a). The error percentage is then 3.6%, which is close to the laser fluence fluctuation (4%). After normalization of the transient history with respect to the peak temperature and filtering, we plotted the normalized error as a function of time and found it to be on the order of 1%(Fig.7b).

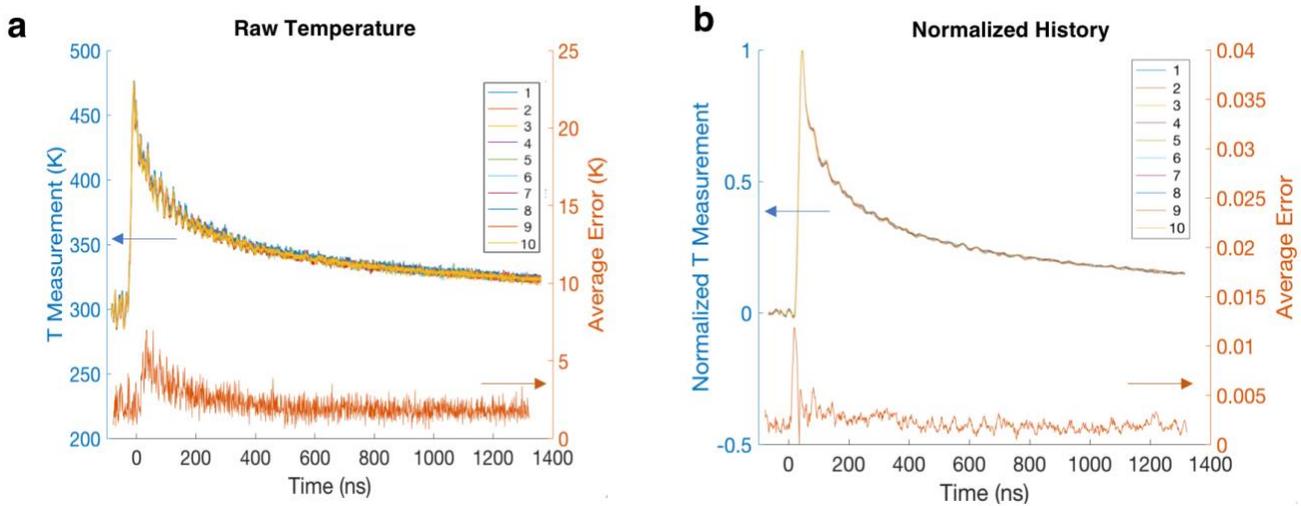

**Figure 7 Overlay of multiple measurements and average error for raw temperature and normalized temperature history.** (**a**) The raw temperature evolution and error, (**b**) the normalized and filtered temperature history and error. The sensor is on silica and insulated with 7nm ALD alumina.

## 4. Overall Validation

Below we describe the experimental verification of the sensor's measurement through a nanosecond transient process with a known reference temperature. The spatial and temporal accuracy as well as the system accuracy are therefore confirmed. We utilize the melting points of the amorphous and crystalline silicon film as known reference temperatures. Thanks to prior laser crystallization studies[22,27], the melting points have been verified as essentially constant in the nanosecond time regime. 40nm of amorphous silicon thin film is deposited through PECVD method on top of 35nm silicon nitride insulation layer due to its high mechanical strength against laser-induced mechanical shock. Lastly, the sample is irradiated through a 13ns pulsed laser (New Wave Polaris II, 532nm). Though the film thickness is smaller than three times the penetration depth, with multi-layer thin film interferences[28], we calculated the absorption of the laser energy that is confined in mainly in a-Si film layer (Fig. 8b, ii).

The laser-induced melting is clearly captured in the plot of peak temperature against the laser pulse fluence (Fig. 8a). For a given sample, we gradually increase the fluence and simultaneously measure the transient temperature evolution. The record the peak temperature is the averaged value in the 10 pulses at given fluence. The probed peak temperature first increased linearly with incident fluence, corresponding to the stage where no phase change is involved. Consequently, the probed temperature rise follows the relation, $\Delta T = Q/C_p$, where Q is the deposited pulse energy on the sensor and $C_p$ is the specific heat. However, at 1.5 J/cm$^2$ the measured temperature deviated from the linear correlation and the deviation was enlarged with increased incident energy. As the laser profiles are Gaussian, we assume the gradual deviation indicates the initiation of melting in the center region and subsequent expansion across the irradiated area. Hence 1.5 J/cm$^2$ is considered to be the onset of the phase transformation. After reaching 2.5 J/cm$^2$, the temperature plateaued and soon started to drop. This stage was interpreted as the completion of full melting and crystallization. In the experiments, we started to reduce the fluence, and the temperature dropped to a level lower than before, exhibiting a hysteresis in the temperature vs. fluence curve, which is visually guided by the blue arrows. The hysteresis is caused by the fact that an increased amount of crystalline silicon in the laser-irradiated area reduced the absorption of the incident wave, leading to the drop in temperature increase. In support of this argument, a multi-layer thin film interference model[28] was solved and the absorption was predicted to drop with increased crystallinity (Fig.8 bii). In the optical image (Fig. 8b), the color of the irradiated area has changed, indicating a local refractive index change caused by the phase transformation. Furthermore, we measured the reflectivity at different locations. Pristine regions had a lower reflection compared to the crystallized region, lending credit to the optical simulation and proposed interpretation.

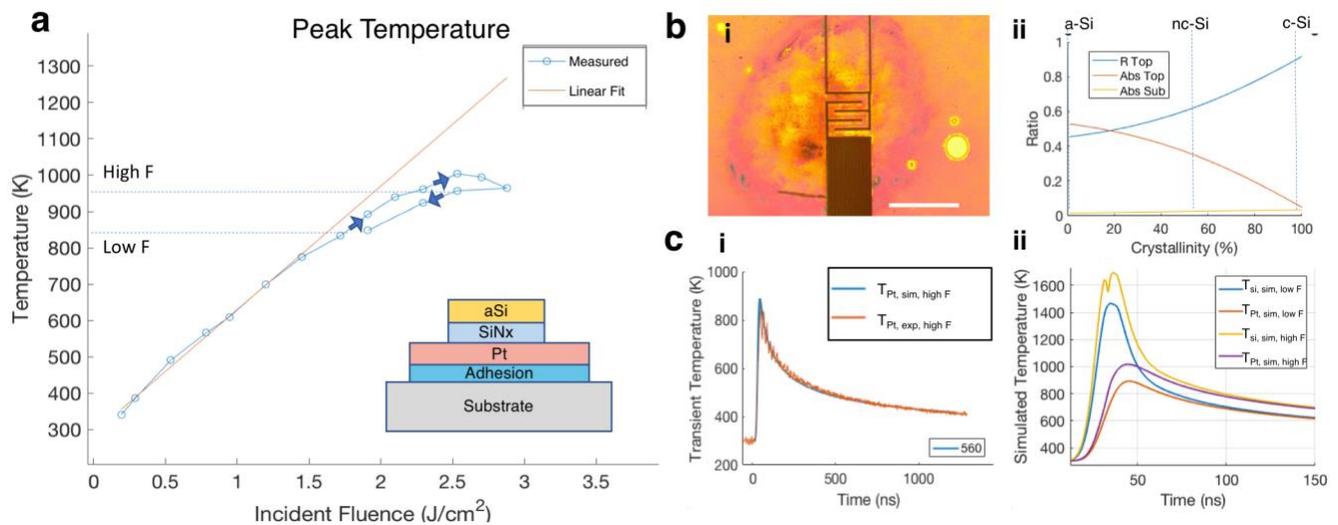

**Figure 8 Experimental validation of the temperature measurement magnitude and temporal evolution**. (**a**) Probed peak temperature against the incident fluence. A linear fit is applied before phase transformation and ablation. The inset is the tested sample geometry. (**b**) (i)Bright field images of laser irradiated 40nm amorphous silicon film on silicon nitride insulating layer. (ii) The multi-layer thin film analysis of the reflection and absorption among layers. Top stands for a-Si film and sub stands for the Pt

layer. (**c**) (i)Experimental and simulated temperature evolution of the Pt layer under insulation and silicon films with a fluence of 2 J/cm$^2$. (ii) Simulated temperature evolution of silicon top surface and Pt layer under High and Low fluence threshold indicated in (a).

The maximum temperature and temporal evolution were confirmed in conjunction with 1D heat transfer simulation. We first verified the agreement of the temporal evolution of the sensor temperature with our modeling. Moreover, we set the modeled incident fluence, F, to ensure the modeled Pt layer maximum temperature matched with the probed maximum temperature at both low F (on-set of amorphous silicon crystallization)and high F (crystalline silicon crystallization). Note that the spatial heterogeneity will cause reduced probe temperature, and the difference for the current sample is estimated less than 50K. Hence, we intentionally compensated 50K to the probed temperature as the input for modeling. The matched model was further used to predict the top layer temperatures under two incident fluences, which is plotted together with the sensor temperature at Fig. 8c, ii. At low F, the top layer reached an amorphous melting point(1485K) and at high F, the crystalline silicon melting point(1685K), perfectly aligned with the observations and physical interpretations of Fig. 8a. Hence, we experimentally confirmed the maximum temperature and temporal evolution of the sensor given the proper accounting of the effects of spatial non-homogeneity. The uncertainty due to the non-homogeneity effects, is within 25K.

## 5. Application: Probing heat transfer among ultrathin films

The peak of transient temperature evolution will be significantly affected by the insulation layer or top layer, which implies the nanosecond-resolved sensing can be used for thin film thermal conductivity study. As a proof-of-concept experiment, we measured the initial part of the laser-induced temperature for samples with 7nm ALD Alumina or 45nm PECVD oxide insulation layers. E-beam evaporation is used for depositing 50nm of amorphous Ge, which is over three times of the optical penetration depth. With a thinner insulation layer, the temperature rise time is shorter and the maximum temperature is higher due to the lower thermal resistance (Fig. 9). The thermal conductivity of PECVD oxide[29] and ALD alumina[30] is set as 1W/mK as suggested by references. The laser fluence is the only tuning parameter and the same fluence is applied to both simulations. Slight differences in the maximum temperature can be caused by the laser energy fluctuation as well as thin-film interference effect on the reflectivity of the sensor. Similar to the Laser Flash Analysis(LFA), the sensor can be readily used to probe the thermal conductivity of unknown material or the thickness of a known material. Different from LFA that only measures a homogeneous bulk sample, the proposed method provides a laterally microscale and vertically nanoscale measurement on heterogeneous multilayer samples. It is also conceivable that layouts of multiple sensors embedded at different depths in microstructured target specimens could provide 3D transient temperature maps under varying thermal loading configurations.

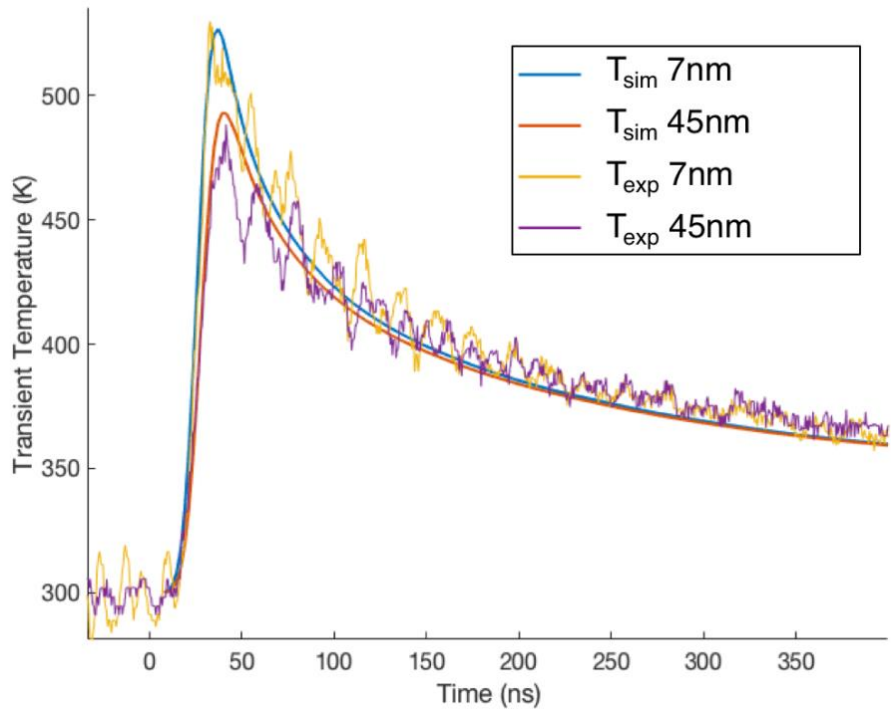

**Figure 9 Experimental and simulated temperature evolution on samples with different insulation layers.** All the samples are fabricated on silica substrates. The legends indicate the thickness of the insulation layer.

# Conclusion

In the current study, we presented the systematic design, optimization and implementation of nanosecond resolved microscale Pt thin-film sensors. The error sources were analyzed and optimized during the development of the sensor. Coupled thermal and electrical simulations quantified the spatial resolution and accuracy imposed by the Gaussian beam and serpentine structure. Regarding the temporal resolution, we characterized the sensor rise time as 9 ns via launching picosecond laser irradiation as a heat impulse input. The temporal evolution of the sensor measurements agree well with simulations on different film structures and substrates. The absolute temperature magnitude was evaluated against amorphous silicon film melting and the error was shown to be below 25K. After correcting for the laser energy fluctuation, the temperature measurement attained fluctuation less than 0.5%. Lastly, we demonstrated the sensor's potential applications for thin film thermal conduction with nanometer sensitivity, which will be three orders of magnitude more sensitive compared to Laser Flash Analysis.

## Acknowledgment

The authors would like to thank Yang Xia in the Department of Applied Science and Technology and Luya Zhang in the Department of Electrical Engineering and Computer Science for helpful discussion on the electrical circuit. The nanofabrication was carried out at the Marvell Nanofabrication Laboratory and the California Institute of Quantitative Bioscience (QB3) of UC Berkeley. The authors thank Chris Zhao from Novellution Technologies, Inc. for helpful discussion on the fabrications. The work received financial support from Lam Research Corp..